\documentclass[journal = apchd5, manuscript = article]{achemso}
\usepackage[english]{babel}
\usepackage{amsmath,amssymb,bm}
\usepackage[utf8]{inputenc}
\usepackage{natbib}
\usepackage[dvipsnames]{xcolor}
\usepackage{url}
\usepackage{balance}
\usepackage{textcomp}
\usepackage{array}
\usepackage{graphicx}
\usepackage{upgreek}
\usepackage{xfrac}
\usepackage{soul}
\usepackage{bm}

\usepackage[bookmarks=true,
            bookmarksnumbered=false,
            bookmarksopen=true,
            breaklinks=true,
            pdfborder={0 0 0},
            final=true,
            backref=none,
            colorlinks=true,
            linkcolor=blue,
            menucolor=blue,
            citecolor=blue,
            urlcolor=blue]{hyperref}

\renewcommand{\mu}{\upmu}
\newcommand{\um}{\mu\textrm{m}}			
\newcommand{\ium}{\mu\textrm{m}^{-1}} 	

\author{M.~Sich}
\affiliation{Department of Physics and Astronomy, The University of Sheffield, Sheffield, S3~7RH, United Kingdom}

\author{L.~E.~Tapia-Rodriguez}
\affiliation{Department of Physics and Astronomy, The University of Sheffield, Sheffield, S3~7RH, United Kingdom}

\author{H.~Sigurdsson}
\affiliation{Science Institute, University of Iceland, Dunhagi-3, IS-107 Reykjavik, Iceland}

\author{P.~M.~Walker}
\affiliation{Department of Physics and Astronomy, The University of Sheffield, Sheffield, S3~7RH, United Kingdom}

\author{E.~Clarke}
\affiliation{EPSRC National Epitaxy Facility, The University of Sheffield, Sheffield, S1~4DE, United Kingdom}

\author{I.~A.~Shelykh}
\affiliation{Department of Nanophotonics and Metamaterials, ITMO University, St. Petersburg, 197101, Russia}
\affiliation{Science Institute, University of Iceland, Dunhagi-3, IS-107 Reykjavik, Iceland}

\author{B.~Royall}
\affiliation{Department of Physics and Astronomy, The University of Sheffield, Sheffield, S3~7RH, United Kingdom}
\affiliation{Huawei, B55 Adastral Park, Ipswich, IP5~3RE, United Kingdom}

\author{E.~S.~Sedov}
\affiliation{School of Physics and Astronomy, University of Southampton, Southampton, SO17 1NJ, United Kingdom}
\affiliation{Vladimir State University, Gorky st. 87, Vladimir, 600000, Russia}

\author{A.~V.~Kavokin}
\affiliation{School of Physics and Astronomy, University of Southampton, Southampton, SO17 1NJ, United Kingdom}
\affiliation{CNR-SPIN, Viale del Politecnico 1, I-00133, Rome, Italy}
\affiliation{Spin Optics Laboratory, St. Petersburg State University, Ul’anovskaya 1, Peterhof, St. Petersburg, 198504, Russia}

\author{D.~V.~Skryabin}
\affiliation{Department of Physics, University of Bath, Bath, BA2 7AY, United Kingdom}
\affiliation{Department of Nanophotonics and Metamaterials, ITMO University, St. Petersburg, 197101, Russia}

\author{M.~S.~Skolnick}
\affiliation{Department of Physics and Astronomy, The University of Sheffield, Sheffield, S3~7RH, United Kingdom}
\affiliation{Department of Nanophotonics and Metamaterials, ITMO University, St. Petersburg, 197101, Russia}

\author{D.~N.~Krizhanovskii}
\email{d.krizhanovskii@sheffield.ac.uk}
\affiliation{Department of Physics and Astronomy, The University of Sheffield, Sheffield, S3~7RH, United Kingdom}
\affiliation{Department of Nanophotonics and Metamaterials, ITMO University, St. Petersburg, 197101, Russia}

\title{Spin domains in one-dimensional conservative polariton solitons}

\makeindex
\begin{document}
\date{\today}

\begin{abstract}
We report stable orthogonally polarised domains in high-density polariton solitons propagating in a semiconductor microcavity wire. This effect arises from spin dependent polariton-polariton interactions and pump-induced imbalance of polariton spin populations. The interactions result in an effective magnetic field acting on polariton spin across the soliton profile, leading to the formation of polarisation domains. Our experimental findings are in excellent agreement with theoretical modelling taking into account these effects.
\end{abstract}

Temporal and/or spatial domains of coupled multiple dipoles play a significant role in the properties of various physical systems.
Magnetic domains in (anti)ferromagnets have been widely studied up to date and are utilised in modern memory devices (hard drives). Domain formation with electric dipoles has also been observed in atomic spinor Bose-Einstein condensates~\cite{Stenger1998}. In optics, polarisation domain formation is governed by modulation instability with the resultant separation of adjacent domains by a domain wall, a topological defect closely linked to soliton formation~\cite{Sheppard1994,PhysRevE.71.066209,Zakharov1970}. While scalar nonlinear effects related to solitons~\cite{OptSolBook,DissipSolBook} (or optical supercontinuum~\cite{Skryabin2010}) have been studied extensively, little attention has been given to the spatiotemporal evolution of the polarisation (or spin) degree of freedom in vectorial optical structures. Only recently, time-localised polarisation rotations or polarisation domain walls have been reported for light travelling in nonlinear optical fibres~\cite{Gilles2017} and for dissipative solitons in vertical-cavity surface-emitting lasers~\cite{Marconi2015}.

In this paper we study spatiotemporal polarisation domains in a system of exciton-polaritons in semiconductor microcavities. In such highly nonlinear systems observation of a variety of quantum fluid phenomena~\cite{Gippius2007, Walker2015}, Bose-Einstein condensation~\cite{bookBEC}, Berezinskii-Kostelitz-Thouless phases, superfluidity, dark and bright solitons~\cite{Sich2016} have all been reported.

Polaritons are characterised by two possible spin projections on the structure growth axis, which correspond to two opposite circular polarisations~\cite{Shelykh2010}.  Efficient external control of the polariton spin makes microcavity based structures promising building blocks for \textit{spin-optronic} devices, \textit{i.e.} optical equivalents of spintronic devices~\cite{Shelykh2004}. Moreover, due to the exchange terms that dominate the exciton-exciton scattering~\cite{Ciuti1998}, polariton-polariton interactions are strongly spin anisotropic: polaritons with same spin projections strongly repel each other, while polaritons with parallel spins interact more weakly and sometimes even attract each other~\cite{Wouters2007,Vladimirova_2010}. The interplay between spin dynamics and polariton-polariton interactions leads to a wide variety of nonlinear polarisation phenomena in microcavities. This includes spin switching~\cite{Amo2010}, spin-selective filtering~\cite{Gao_SpinFilt2015}, polarisation dependent stability of dissipative solitons~\cite{Sich2014}, optical analogues of magnetically ordered states~\cite{Ohadi2017,Sigurdsson_2017}, dark half solitons~\cite{Hivet2012}, and spin and half vortices~\cite{Rubo2007,Dominici_2018}.

Here we demonstrate formation of polariton spin domains within high-density wavepackets evolving into conservative soliton(s) in a quasi-1D system. Experimentally we resolve the full Stokes polarisation vector of the propagating wavepackets under a wide range of excitation powers, and develop a theoretical model which we validate through numerical simulations, qualitatively reproducing the experimental results. The
experiments were performed in a microcavity wire (MCW): 5-$\um$ wide and 1-mm long mesas etched from planar microcavities. Thanks to the 1D nature of polariton wires and the long polariton lifetimes ($\sim 30$ ps), we could reach polariton densities sufficient for formation of solitons~\cite{Skryabin2017}. At low excitation powers we observe polarisation precession caused by a linear in-plane effective magnetic field inside the sample. At higher excitation powers the high polariton density and an imbalance of polariton spin populations lead to an extra out-of-plane magnetic field. This nonlinear field causes, firstly, fast polarisation oscillations, and then the formation of spin domains.

\begin{figure}
\centering
\includegraphics[width=\columnwidth]{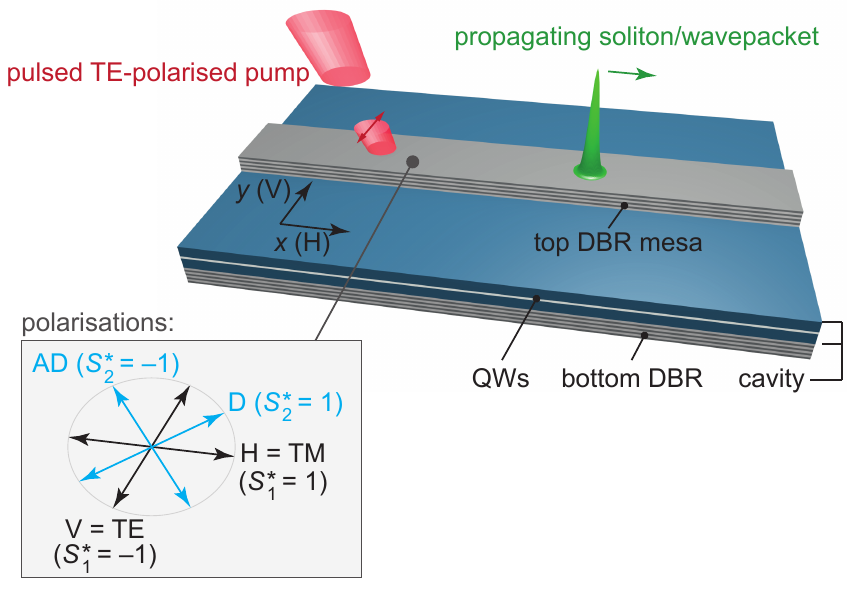}
\caption{Scheme of the experiment. The partially etched top DBR mesa aligned along the longitudinal axis, $x$. Bottom-left inset shows linear polarisations mapped onto the laboratory frame relative to the sample position: horizontal (H), vertical (V), diagonal (D), and anti-diagonal (AD); and their corresponding Stokes vector $(\mathbf{S}^*)$ component values.}
\label{fig:intro}
\end{figure}

\section{Results}
We performed our experiments on a $\sfrac{3\uplambda}{2}$ microcavity  composed of 3 embedded InGaAs quantum wells (10~nm thick, 4\% Indium) and GaAs/AlGaAs (85\% Al) distributed Bragg mirrors with 26 (23) repeats on the bottom (top) mirror. The Rabi splitting and the polariton lifetime are $\simeq$~4.12~meV and \({\simeq 30}\)~ps in this sample, which was previously described in Refs.~\citenum{Tinkler2015,Skryabin2017}.  The detuning between the exciton and the photon modes is $\simeq-2$~meV. The top mirror was partially etched (down to the last few layers of the top DBR) defining $1000~\um$ long mesas (MCWs) of different widths. In all our measurements we used the same $5~\um$-wide MCW as in Ref.~\citenum{Skryabin2017}.

We excited the sample using a TE polarised pulsed (\(\simeq5\)~ps full width at half-maxima, FWHM) laser quasi-resonant with the lower polariton branch corresponding to the ground MCW mode. The angle of incidence was $k_x\simeq2.2~\ium$, at which the effective polariton mass is negative, favouring formation of solitons \cite{Skryabin2017}. We employed transmission geometry, where we applied the excitation beam on one side of the cavity (bottom) and collected emission on the opposite side (top) to avoid the reflected pump beam saturating our detectors. We then changed the laser power and measured the full Stokes vector of the emitted light as a function of time and propagation coordinate, $x$, for each excitation power. In the experiment we separately detected emission intensity in six polarisations: horizontal ($I_h$), vertical ($I_v$), diagonal ($I_d$), anti-diagonal ($I_{ad}$), right-hand ($I_{\sigma^+}$), and left-hand ($I_{\sigma^-}$) circularly polarised. In the circular polarisation basis of the cavity light $\left\{ \psi_+;\psi_- \right\}$, the total intensity is written $S_0 = |\psi_+ |^2  + |\psi_-|^2$ and the Stokes vector (normalised to unity) $\mathbf{S^\ast}=(S_1^\ast,S_2^\ast,S_3^\ast)$ is given by
\begin{eqnarray}
S_1^\ast &=& \frac{ \psi_+ \psi_-^* + \psi_+^* \psi_-  }{|\psi_+ |^2  + |\psi_-|^2} = \frac{I_h-I_v}{I_h+I_v},\label{eq:S1} \\
S_2^\ast &=& i\frac{ \psi_+ \psi_-^* - \psi_+^* \psi_-  }{|\psi_+ |^2  + |\psi_-|^2}= \frac{I_d-I_{ad}}{I_d+I_{ad}},\label{eq:S2} \\
S_3^\ast &=& \frac{ |\psi_+ |^2  - |\psi_-|^2}{|\psi_+ |^2  + |\psi_-|^2} = \frac{I_{\sigma^+}-I_{\sigma^-}}{I_{\sigma^+}+I_{\sigma^-}}.\label{eq:S3}
\end{eqnarray}

\begin{figure}
\centering
\includegraphics[width=\columnwidth]{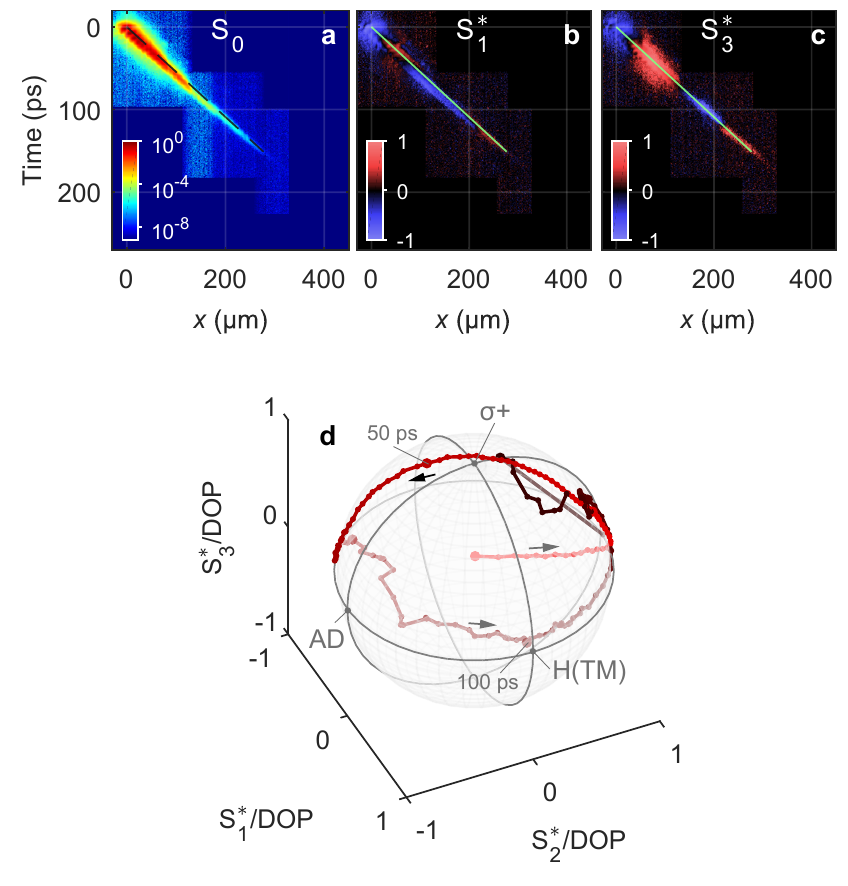}
\caption{$P=87~\mu$W. \textbf{a} Total emission intensity ($S_0$ Stokes component) of a soliton as a function of coordinate x and time. \textbf{b, c} Degree of polarisation of the emission in linear ($S_1^\ast$), and circular ($S_3^\ast$) polarisation bases. \textbf{d} Experimental evolution of the tip of the polariton Stokes vector on the surface of the Poincar\'{e} sphere as a function of time. The length of the experimental Stokes vector is normalised to unity. The dimmer (brighter) traces correspond to the evolution on the Stokes vector on the back (front) surface of the sphere. The North and the South poles of sphere correspond to $\sigma^+$ and $\sigma^-$ circular polarisations, whereas the points on the equator correspond to the linear polarisations. The direction of the Stokes vector is constructed by measuring the values of the Stokes components at different times at the spatial points of the soliton profile shown by the green lines in panels \textbf{b} and \textbf{c}. Data on a-c is taken at $y=0\pm0.5~\um$.
}
\label{fig:exp1}
\end{figure}

At the lowest excitation power, $P=13~\mu$W, polariton nonlinearity is very weak and the wavepacket propagates in the linear regime, experiencing dispersive spreading (Fig.~4 of the SM). At higher power, $P=87~\mu$W, a soliton is formed, characterised by non-spreading propagation~\cite{Sich2012,Skryabin2017} as shown in Fig.~\ref{fig:exp1}a. Both circular and linear polarisation degrees experience oscillations between negative and positive values with time as shown in the $S_1^\ast$ and $S_3^\ast$ components (Figs.~\ref{fig:exp1}b,c) and the $S_2^\ast$ component (Fig.~3 of the SM). We note that even though the pump beam is TE-polarised, the polarisation of the emission at time $t=0$ has some diagonal and circular components, likely due to birefringence in the substrate and the influence of the edges of the MC wire on the polarisation of the pump field before it couples to the polariton field inside the wire. 

The polarisation beats observed in Figs.~\ref{fig:exp1}b,c correspond to motion of the Stokes vector $\mathbf{S}^*$ (which is also referred to as the polariton pseudospin or spin) around the unit Poincare sphere, as shown in Fig.~\ref{fig:exp1}d where we plot the trajectory that the tip of the normalised Stokes vector follows as the soliton propagates. The direction of the Stokes vector is constructed by measuring the values of the Stokes components at different times at the spatial points of the soliton profile shown by the green lines in in Fig.~\ref{fig:exp1}a,b, and c. The Stokes vector clearly precesses around the sphere as the soliton propagates. We note that at the lowest excitation power (P=13~$\mu$W) the polariton wavepacket does not propagate long distances due to fast spreading and the onset of the polarisation beats and Stokes vector precession are barely visible (see Fig.~3 of the SM). 

Such motion of the Stokes vector around the Poincare sphere can be described mathematically as a precession of the polariton pseudospin around a time-varying effective magnetic field~\cite{Kavokin2004}. Physically, this effective magnetic field arises from three mechanisms: TE-TM (V-H in the laboratory frame) splitting of polaritons propagating with nonzero momenta~\cite{Langbein2007,Hivet2012}; strain, electronic anisotropy or anisotropy related to crystallographic lattices ~\cite{Amo2009,Glazov2010} inducing splitting between D and AD polarised components; and polariton-polariton interactions. The last mechanism is weak at small powers ($P=87$ $\mu$W) , but starts playing a critical role at higher polariton densities ($P> 0.95$ mW) as we discuss below in the modelling section. The observed precession period of $T\simeq60$~ps at $P=87$ $\mu$W (Figs.~\ref{fig:exp1}b-d) corresponds to approximately $\Delta E_{\textrm{eff}} = 2 \hbar/T\simeq 22~\mu$eV energy splitting induced by the effective magnetic field.

\begin{figure}
\centering
\includegraphics[width=\columnwidth]{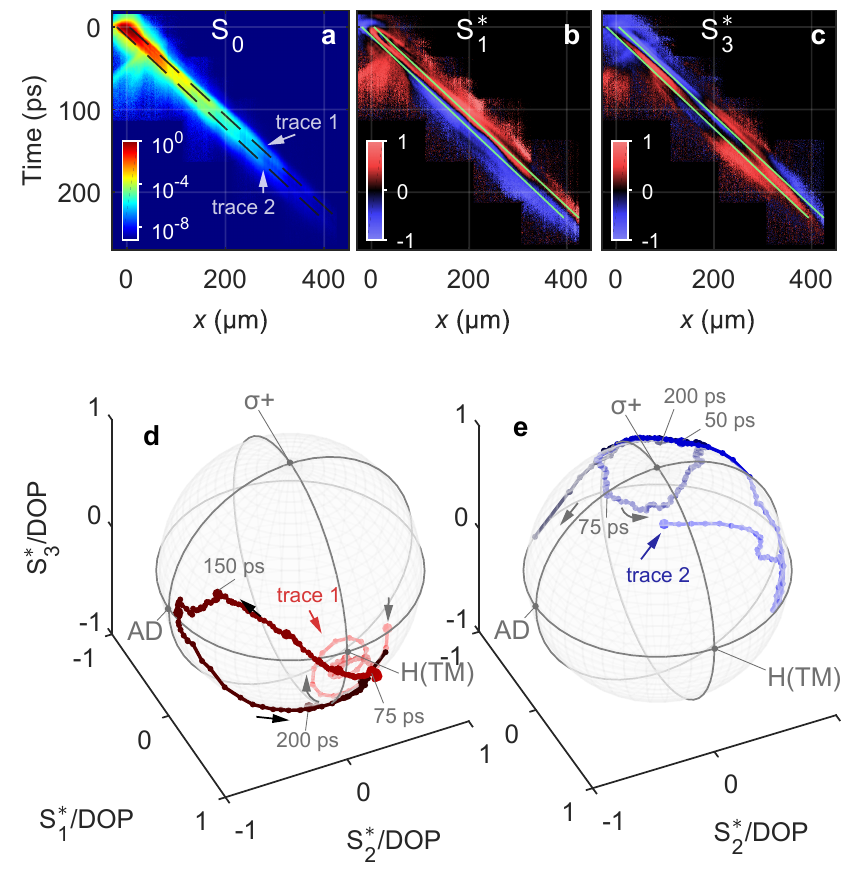}
\caption{
$P=0.95~$mW. \textbf{a} Total emission intensity ($S_0$ Stokes component) of a soliton doublet as a function of coordinate x and time. \textbf{b, c} Degree of polarisation of the emission in linear ($S_1^\ast$), and circular ($S_3^\ast$) polarisation bases. \textbf{d, e} Experimental evolution of the tip of the polariton Stokes vector on the surface of the Poincar\'{e} sphere as a function of time measured at the spatial points of the soliton profiles shown by the black dashed/solid green lines in panels \textbf{a, b} and \textbf{c} for traces 1 and 2 respectively in panels \textbf{d} and \textbf{e}.  The dimmer (brighter) traces correspond to the evolution on the Stokes vector on the back (front) surface of the sphere. The Stokes vectors are normalised to unity. Data on a-c is taken at $y=0\pm0.5~\um$.}
\label{fig:exp2}
\end{figure}

At the intermediate power, $P=0.95$~mW (see Fig.~\ref{fig:exp2}), we observe that the excited polariton wavepacket emits Cherenkov radiation at $t\sim 30$ ps (see Ref.~\citenum{Skryabin2017}) and further evolves into a soliton doublet at $t\sim 50$ ps. 
At this pump power the polarisation dynamics exhibit a drastic change.
Formation of a soliton doublet ($t\sim 50$~ps) is accompanied 
by the establishment of two almost orthogonally-polarised spatially separated polarisation domains associated with each of the soliton spatial components. The domain formation is most pronounced in the $S_1^\ast$ component (H-V linear polarisation basis).
Figs.~\ref{fig:exp2}d and e show the evolution of the normalised Stokes vectors on the Poincare surface, with one soliton of the doublet in each panel. The Stokes vectors are taken at the positions depicted by the dashed black lines (labelled as "trace1" and "trace 2" ) in Fig.~\ref{fig:exp2}a, each going though the components of the soliton doublet. For "trace1" very fast precession of the soliton Stokes vector around the 'South' pole of the Poincar\'{e} sphere is observed during the first 50~ps (bright red trace on Fig.~\ref{fig:exp2}d) with a period of $T\simeq10$~ps,  corresponding to an out-of-plane increased effective magnetic field due to the spin-dependent polariton nonlinearity which induces an energy splitting $\Delta E_{\textrm{eff}} \simeq 130~\mu$eV. 
As the polariton density, and hence nonlinearity, are reduced at times $t>50$~ps the precession occurs at slower speed with a period of $\sim 100$-$125$~ps. In this time range the Stokes vectors taken at positions of "trace 1" and "trace 2" precess in the south and north Poincare hemispheres respectively remaining almost orthogonal. 

\begin{figure}
\centering
\includegraphics[width=\columnwidth]{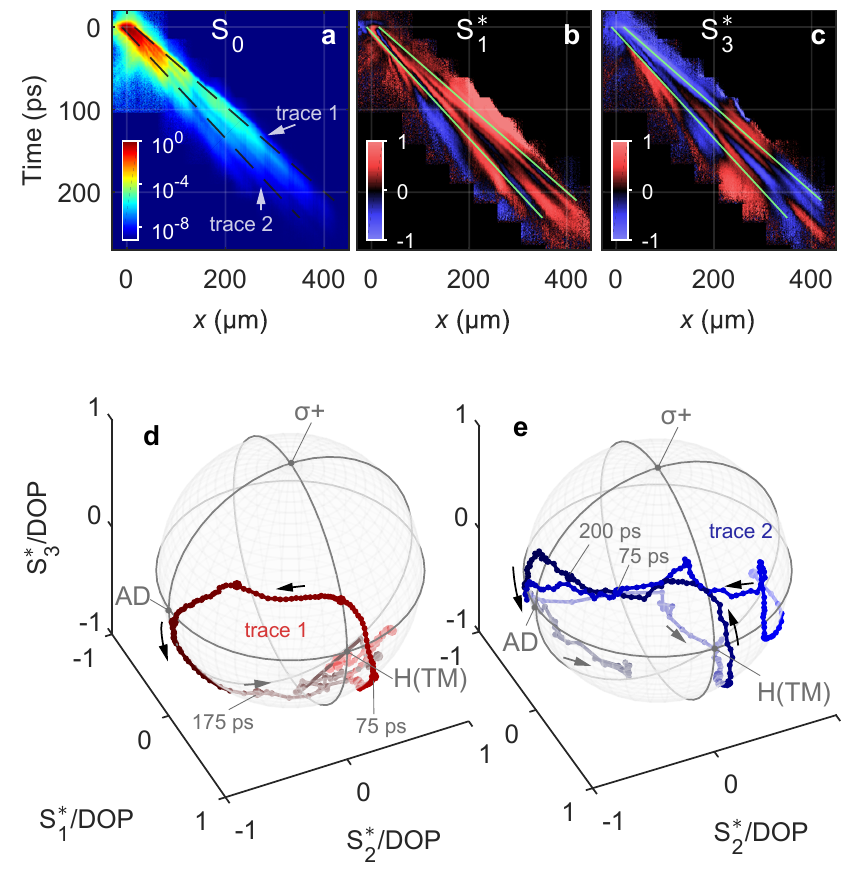}
\caption{$P=3.1~$mW. \textbf{a} Total emission intensity ($S_0$ Stokes component). \textbf{b, c} Degree of polarisation of the emission in linear ($S_1^\ast$), and circular ($S_3^\ast$) polarisation bases. \textbf{d, e} Experimental evolution of the tip of the polariton Stokes vector on the surface of the Poincar\'{e} sphere as a function of time for traces 1 and 2 respectively. The Stokes vector was measured at the spatial points of the polariton wavepacket profiles shown by the black dashed (solid green) lines in panel \textbf{a} (\textbf{b,c}).  The dimmer (brighter) traces correspond to the evolution on the Stokes vector on the back (front) surface of the sphere. The Stokes vectors are normalised to unity. Data on a-c is taken at $y=0\pm0.5~\um$.}
\label{fig:exp3}
\end{figure}

At the highest pump power, $P=3.1$~mW (see Fig.~\ref{fig:exp3}), a further onset of cascaded polariton-polariton scattering leads to the occupation of states on the lower polariton branch at energies below the pump energy in a process resembling the optical continuum generation as observed and discussed in Ref.~\citenum{Skryabin2017} for the same MC wire polariton system. Since in such a process polariton relaxation results in a significant occupation of the states described by positive effective mass~\cite{Skryabin2017}, this  results in both coexisting solitons and spreading dispersive polariton wavepacket with time. At such high polariton density the nonlinearity results in more complex spatio-temporal polarisation dynamics, which is shown in Fig.~\ref{fig:exp3}. 
The domains with negative and positive circular polarisation degree remain for the first $\sim50$-$70$~ps as seen for $S_3^\ast$ component in Fig~\ref{fig:exp3}c. However, at later times when the nonlineairty becomes weaker the wavepacket breaks into a set of domains with different polarisations, the Stokes vectors of which evolve in time in a complex manner as shown in  Figs.~\ref{fig:exp3}d) and e). This temporal evolution originates from the time- and space-dependent out-of-plane  effective magnetic field due to the spin-dependent polariton nonlinearity.

%
%
%

\section{Model}
In order to describe the polarisation dynamics of the nonlinear polariton wavepacket and understand the physical mechanisms responsible for the observed behaviour we work with the coupled spinor macroscopic cavity photon field $\bm{\psi} = (\psi_+,\psi_-)^T$ and exciton wavefunction $\bm{\chi} = (\chi_+, \chi_-)^T$ written in the circular polarisation basis~\cite{Carusotto_PRL2004}. The TE-TM splitting of the photon modes is described by the operator $\Sigma_{\mathbf{k},\pm} = \beta \left( k_{x} \pm i k_{y} \right) ^2$ written in reciprocal space~\cite{Panzarini1999, Kavokin_PRL2005} and characterised by a splitting constant $\beta$.

Propagation distance of polaritons, as well as the spatial scales of the observed effects, significantly exceed the width of the channel. This allows us to take advantage of the 1D nature of the microwire in order to simplify the formalism. Let us first consider an infinite rectangular potential well width~$w_{y}$ where we assume the photon wavefunction to be in the form $\psi _{\pm} (t,\mathbf{r}) = \sqrt{2/w_{y}} \cos (\pi y / w_{y}) \exp{(- i \varepsilon _{y} t / \hbar)} \psi_{\pm}(t,x)$. The set of coupled equations for the 1D polariton envelope $\psi_{\pm}(t,x)$ propagating in the $x$-direction can be written:
\begin{align} \notag 
i \hbar \frac{\partial \psi_\pm}{\partial t} & =  \left[ - \frac{\hbar^2 }{2 m}\frac{\partial^2}{\partial x^2}+ \Delta - i  \frac{\hbar \Gamma}{2} \right] \psi_\pm + (\Sigma_x  \mp i \delta) \psi_\mp \\
 & + \frac{\hbar \Omega}{2} \chi_\pm + E_\pm e^{-i(\omega_p t -k_p x -i x^2/2w_x^2 -i t^2/2w_t^2)}, \label{eq:GPEs} \\
i \hbar \frac{\partial \chi_\pm}{\partial t}  = & \left[\alpha_1 |\chi_\pm|^2 + \alpha_2 |\chi_\mp|^2  - i \frac{\hbar \Gamma_\chi}{2} \right]\chi + \frac{\hbar \Omega}{2}\psi_\pm. \label{eq:GPEs2} 
\end{align}
Excitons possess a much larger effective mass than cavity photons and their dispersion can be safely approximated as flat compared to the photon dispersion. Here, $m$ is the effective mass of cavity photons; $\Delta$ the exciton-photon detuning; $\Gamma$ is cavity photon decay rates corresponding to photons leaking from the cavity; $\Gamma_\chi$ is the exciton decay rate corresponding to nonradiative dephasing processes; $\Sigma_x = -\beta \left(\partial_x^2 + \pi ^2 / w_y^2 \right)$ is the real space TE-TM operator along the microwire; $\delta$ describes a fixed splitting in the diagonal polarisations which arises from the optical and electronic anisotropy in a microcavity~\cite{Amo2009,Glazov2010}; and $\hbar\Omega$ is the Rabi splitting giving rise to the exciton-polariton eigenmodes of the system. The parameters $\alpha _1$ and $\alpha _2$ are the interaction constants in the triplet configuration (parallel spins) and in the singlet configuration (opposite spins), respectively. The last term in Eq.~\eqref{eq:GPEs} describes the resonant optical pumping of the lower branch polaritons where $E_\pm$ describes the pump pulse amplitude and phase; $w_x$ and $w_t$ are the spatial and temporal pulse width respectively; $\hbar \omega_p$ is the pump energy; and $k_p$ the pulse wavevector along the wire.

It is instructive to describe the polarisation effects in the microwire in terms of an effective magnetic field which acts on the Stokes vector~\cite{PhysRevLett92017401}. The magnetic field vector can be written, in units of energy, as $\mathbf{\Omega} = (\Omega_x, \Omega_y,\Omega_z)$ where $\Omega_x = \beta(k_x^2 - \pi^2/w_y^2)$, $\Omega_y = \delta$, and $\Omega_z = (\alpha_1-\alpha_2)(|\chi_+|^2 - |\chi_-|^2)$. The last term is a consequence of the nonlinear interactions between polaritons giving rise to effective Zeeman splitting when the spin populations are imbalanced.

For modelling, we take the following values of the parameters: $\beta = 12 \, \mu \text{eV} \, \mu\text{m} ^{2}$; $\delta _l = 20 \, \mu\mathrm{eV}$; $\alpha_1 = 2 \, \mu \text{eV} \, \mu \text{m}$; $\alpha_2 = -0.1\alpha_1$, $\Gamma = 1/30\,\text{ps}^{-1}$, $\Gamma_\chi = 2 \Gamma$; $m = 5 \cdot 10^{-5} m_{0}$, where $m_0$ is the free electron mass; $\Delta = -2$~meV; $\hbar \Omega = 4.12$~meV; $\hbar \omega_p = -1.3$~meV; $k_p = 2.2 \, \mu\text{m}^{-1}$; $w_x$ and $w_t$ were set to have FWHM of 20 $\mu$m and 5 ps respectively. The initial polarisation of the resonant laser is chosen to fit the experimental results: $E_+ =  E_0$ and $E_- = 2.65 e^{i\phi} E_0$ where $\phi = 1.9$ and $E_0$ denotes the overall excitation amplitude of the beam.

\section{Simulations}
\begin{figure}[t!]
\centering
\includegraphics[width=\columnwidth]{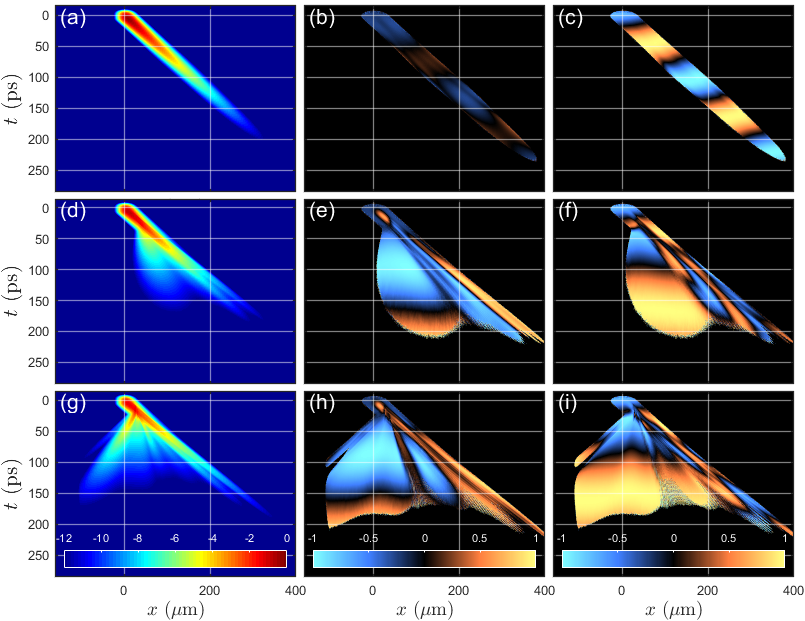}
\caption{
\label{figS0xzL}
\textbf{a,d,g} Total intensity $S_0$ (logarithmic scale) and degrees of linear, $S_1^\ast$ \textbf{b,e,h}, and circular, $S_3^\ast$ \textbf{c,f,i} polarisations. Here $E_0 = \{0.4, \ 1.5, \ 2.0\}$ meV $\mu$m$^{-1/2}$ respectively for top, middle and bottom panel rows.}
\end{figure}

The polarisation dynamics of the resonantly excited polariton pulse are shown in Fig.~\ref{figS0xzL}a-c at low excitation power  where $\Omega_z \ll \Omega_x,\Omega_y$. The total intensity distribution in Fig.~\ref{figS0xzL}a already demonstrates the soliton behaviour as in the experiment (Fig.~\ref{fig:exp1}a). Both the linear (Fig.~\ref{figS0xzL}b) and circular (Fig.~\ref{figS0xzL}c) polarisation components experience harmonic oscillations with frequency $\Omega/\hbar = \sqrt{ \beta \left( k_p^2 - \pi ^2 / w_y^2 \right) + \delta^2}/\hbar$, and shifted in phase relative to each other in line with the experimental observations (Fig.~\ref{fig:exp1}b-c). We note that in the weak pulse regime the effect of TE-TM splitting does not account for the appearance of oscillations in the linear polarisation $S_1^*$ component since the effective magnetic field is oriented in the $x$-direction. The oscillations of $S_1^\ast$ emerge when the additional splitting $\delta$ between the diagonal polarizations is accounted for in the microwire system.

Fig.~\ref{figS0xzL}d-i shows simulation of the nonlinear polarisation dynamics of the polariton pulse for intermediate and high excitation powers. Both the intensity (Fig~\ref{figS0xzL}d,g) and the $S_1^*$ and $S_3^*$ polarisation (Fig.~\ref{figS0xzL}e,h and \ref{figS0xzL}f,i respectively) effects observed experimentally in (Fig.~\ref{fig:exp2}b-c and Fig.~\ref{fig:exp3}b-c) are reproduced in the modelling. The polariton pulse transforms to a soliton doublet at early stages of propagation as in the experiment (Fig.~\ref{fig:exp2}a) and is characterized by long streaks of constant $S_1^*$ polarization as shown in Fig.~\ref{figS0xzL}e,h again as in the experiment (see Fig.~\ref{fig:exp2}b). The reason for this effect is that the $\Omega_z$ component decreases as the polariton intensity decays, becoming small or comparable to the in-plane magnetic field $(\Omega_x,\Omega_y)$ around $\sim50$ ps. Consequently, the rotation of the $S_1^*$ Stokes vector component halts and long streaks are formed in the microwire.

At the low excitation power in Fig.~\ref{figS0xzL}a-c the effective, interaction induced, $z$-magnetic field has its maximum absolute value of $\Omega _z \sim 0.01$ meV at around 5 ps after the pulse. The field is small and the polarisation precession is mainly governed by splitting between linearly polarised components. Increasing the excitation power, the $\Omega _z$ component becomes significant leading to fast polarisation changes. For intermediate pulse energies [Fig.~\ref{figS0xzL}d-f] its value is $\Omega _z \sim 0.45$ meV and $\sim 0.02$ meV at $t \sim 20$ ps and $50$ ps after the pulse respectively. For large pulse energies [Fig.~\ref{figS0xzL}g-i] its value is $\Omega _z \sim 1.6$ meV and $\sim 0.04$ meV at $t \sim 10$ ps and $50$ ps after the pulse respectively.

Cherenkov radiation is also observed before the pulse separation (Figs.~\ref{figS0xzL}d,g). At the highest power complex spatio-temporal dynamics of the polarisation Stokes components (Figs.~\ref{figS0xzL}h-i) resemble those observed in the experiment. (Figs.~\ref{fig:exp3}b-c)

We finally point out that although the polariton nonlinearity plays a crucial role in the spin domain formation,  the formation of solitons in the system is only weakly influenced by the presence of effective magnetic fields from various polarisation splitting mechanisms. We find that approximately the same spatial patterns of solitons appear in the intensity of scalar wavefunctions and the total intensity of the spinor wavefunctions considered here (see Sec.~B in the SM). Thus, the initial formation of the solitons at the pump spot (e.g. the pattern of total intensity summed over all polarisations which leads to non-dispersive propagation) is dominantly a scalar effect. Separately, the nonlinear pseduo-magnetic field leads to the formation of the polarisation domain pattern on top of the solitonic total-intensity pattern.

\section{Conclusion}

We have observed nonlinear spin dynamics of polariton wavepackets in a microcavity wire. At low excitation just above the threshold of soliton formation the polarisation precession during the propagation is caused by an effective magnetic field due to splittings between linearly polarised polariton component. With increasing polariton density an extra pump-induced magnetic field due to anisotropy in the spin-dependent polariton-polariton interactions  results in formation of spatially separated polarisation domains. While soliton formation is largely a scalar effect (not dependent on polarisation), it likely helps to preserve spin domains over the propagation distance due to the non-spreading nature of solitons, hence polarisation domains experience little dispersion.

Polariton wires considered in this work can represent building blocks for future polaritonic devices, such as polariton based spin transistor~\cite{Solnyshkov2009,Johne2010} or soliton-based logic gates~\cite{Cancellieri2015}. In the context of the recent demonstration of bright temporal conservative solitons in such a system~ \cite{Skryabin2017} polarisation domains may be used for the non-binary information encoding and transfer. This observation also opens a possibility for further fundamental studies of the system, such as description of domain wall formation mechanisms.
It would be also interesting to investigate polarisation dynamics of bright solitons in high velocity thick waveguide ($>1 \mu$m) systems~\cite{Shapochkin_2018}, where the splitting between linearly polarised components is comparable to polariton blueshifts. Recently, waveguide polaritons have been also reported in a screening resistant GaN system~\cite{Ciers_2017}, which may potentially work at 300 K paving the way towards polaritonic device applications.

\begin{acknowledgement}
LETR acknowledges support from CONACYT, the Mexican Council for Sciences. HS and IAS acknowledge support by the Research Fund of the University of Iceland, The Icelandic Research Fund, Grant No. 163082-051. This work was supported by megagrant 14.Y26.31.0015 of the ministry of education and science of Russian Federation. IAS acknowledges support from Goszadanie No. 3.2614.2017/4.6. MSS and DNK acknowledge the support from UK EPSRC grant EP/N031776/1.
\end{acknowledgement}

\section{Supplemental Material}
\subsection{A. Pump induced effective magnetic field}
\begin{figure}[t!]
\centering
\includegraphics[width=\columnwidth]{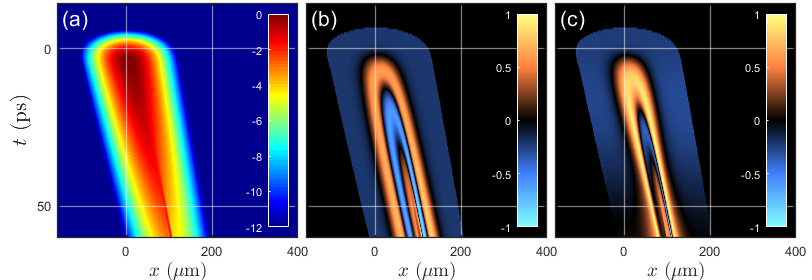}
\caption{
\label{fig.num_supp1}
\textbf{a} Total intensity $S_0$ (logarithmic scale) and \textbf{b,c} linear, $S_1^\ast$ polarisation. Here $E_0 = 2.5$ meV $\mu$m$^{-1/2}$ and FWHM of 80 $\mu$m. In \textbf{b} we set $\beta = \delta  =0$ whereas in \textbf{c} they have the same parameter values as given in the main manuscript.}
\end{figure}

In this section we demonstrate that the formation of polarisation domains in the $S_1^*$ (linear) polarisation appears as a result of a pump-induced, out-of-plane, effective magnetic field. The magnetic field originates from finite ellipticity in the pump beam which creates an imbalance in the two polariton spin populations. The interactions coming from the exciton fraction of the polariton lead to a splitting $\Omega _z (x,t) = (\alpha_1 - \alpha_2) (|\chi_+|^2 - |\chi_-|^2)$. In the absence of scattering processes and soliton formation the exciton wavefunction takes up the Gaussian shape of the pump and we can simply write $\Omega _z (x,t) \approx (\alpha_1 - \alpha_2) (|E_+|^2 - |E_-|^2)\exp{(- \Gamma t-  x^2/2w_x^2)}$. In Fig.~\ref{fig.num_supp1} we show this initial spatio-temporal evolution of $S_1^*$ polarization for intermediate excitation strengths and for a wide pump spot of FWHM 80~$\um$ to bring out the pseudospin precession more clearly. Fig.~\ref{fig.num_supp1}a,b shows the emission intensity and liner polarisation in the absence of TE-TM and diagonal splitting. Fig.~\ref{fig.num_supp1}c is the same as panel b but with non-zero TE-TM and diagonal splitting (same values as in the main manuscript). These results, obtained numerically, are symmetric relative to the centre of the excitation spot -- in contrast to the experimental data. This is likely to be the result of different velocities of TE and TM components in the experiment, and, hence, an asymmetric distribution observed experimentally.

\subsection{B. Formation of solitons}
\begin{figure}[t!]
\centering
\includegraphics[width=\columnwidth]{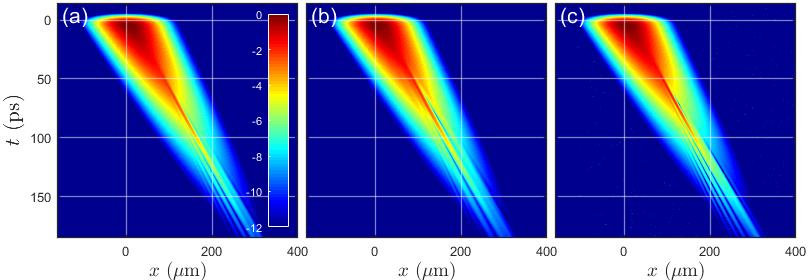}
\caption{
\label{fig.num_supp2}
Total intensity $S_0$ (logarithmic scale) showing soliton streaks. \textbf{a,b} Results of simulating the spinor wavefunction [Eqs.~(4)-(5) main manuscript] without and with TE-TM and diagonal splitting respectively. \textbf{c} Simulation of a scalar wavefunction under the same excitation power.}
\end{figure} 
In this section we investigate the formation of solitons in both spinor and scalar cases. Our results are presented in Fig.~\ref{fig.num_supp2}. In the absence of TE-TM and diagonal splitting we show in Fig.~\ref{fig.num_supp2}a the initial fission of the wavepacket into a soliton train in the spinor picture [Eqs.~(4)-(5) main manuscript]. The resulting pattern matches closely to the resulting pattern calculated using a scalar equation of motion shown in Fig.~\ref{fig.num_supp2}c. 

When both TE-TM and diagonal splittings are present, the solitons diffuse faster and their streaks in the $x$-$t$ plane become more spread (see Fig.~\ref{fig.num_supp2}b). However, the general pattern remains the same and thus the formation of solitons can be attributed to a scalar phenomenon and only weakly influenced by the presence of effective magnetic fields in the MCW.

\bibliography{main}

\end{document}